\documentclass[prd,twocolumn,showpacs,notitlepage,nofootinbib,superscriptaddress]{revtex4-2}
\usepackage{physics}
\usepackage{amsmath}
\usepackage{amssymb}  
\usepackage{graphicx}  
\usepackage{verbatim} 
\usepackage{color}    
\usepackage{subfigure}  
\usepackage{hyperref}  
\usepackage{mathrsfs}
\usepackage{bbold} % Used to cancel terms in equations
\usepackage{cancel} % Used to cancel terms in equations
\usepackage{empheq} % Used for boxing multiline equations
\usepackage{subfigure}

\begin{document}

\title{Implications of  Recent KATRIN Results for Lower-Limits on Neutrino Masses}
\date{\today}
\author{Ephraim Fischbach}
\affiliation{Department of Physics  and Astronomy, Purdue University, West Lafayette, IN 47907, USA}

\author{Dennis E. Krause}
\affiliation{Physics Department, Wabash College, Crawfordsville, IN 47933, USA}
\affiliation{Department of Physics  and Astronomy, Purdue University, West Lafayette, IN 47907, USA}

\author{Quan Le Thien}
\affiliation{Department of Physics, Indiana University, Bloomington, IN 47405, USA}

\author{Carol Scarlett}
\affiliation{Physics Department, Florida A\&M University, Tallahassee, FL 32307, USA}

\begin{abstract}
Recently announced results from the KATRIN collaboration imply an upper bound on the effective electron anti-neutrino mass $m_{\nu_{e}}$, $m_{\nu_{e}}< 0.8~{\rm eV}/c^{2}$.  Here we explore the implications of combining the KATRIN upper bound using a previously inferred lower bound  on the smallest neutrino mass state, $m_{i,{\rm min}}\gtrsim 0.4~{\rm eV}/c^{2}$ implied by the stability of white dwarfs and neutron stars in the presence of long-range many-body neutrino-exchange forces.   By combining  a revised  lower bound estimate with the expected final upper bound from KATRIN,  we find that the available parameter space for $m_{\nu_{e}}$  may be closed completely within the next few years.  We then extend the argument when a single light sterile neutrino flavor is present to set a lower mass limit on sterile neutrinos.

\end{abstract}

\maketitle

Since the discovery of neutrino oscillations \cite{2015 Nobel Prize}, the determination of the masses of neutrinos has been at the forefront of neutrino physics \cite{Otten}.  This problem is especially difficult because the neutrino flavor states, which appear in Standard Model (SM) interaction matrix elements, are superpositions of the  neutrino mass states \cite{Suekane,Bilenky,Giunti,Xing}.  Furthermore, these masses are much less than the masses of any of the other SM particles \cite{PDG 2022}.  Even the mechanism for how neutrinos acquire their masses is not understood \cite{deGouvea}. While cosmological arguments can be used to obtain  stringent limits on the sum of neutrino masses (e.g., $\sum m_{\nu} < 0.13$~eV \cite{Abbott}),  these arguments are heavily model dependent \cite{Otten}.  To date, the most reliable laboratory method of setting limits on neutrino masses has come from examining the endpoint $\beta$-decay spectrum of tritium \cite{Otten}.

It is against this backdrop that the Karlsruhe Tritium Neutrino (KATRIN) experiment collaboration \cite{Aker} has recently published its  most stringent constraint on the effective electron anti-neutrino mass $m_{\nu_{e}}< 0.8~{\rm eV}/c^{2}$ at a 90\% confidence level \cite{KATRIN}.  Here, the effective electron anti-neutrino mass is defined as
\begin{equation}
m_{\nu_{e}}^{2} \equiv \sum_{i= 1}^{3}|U_{ei}|^{2}m_{i}^{2},
\end{equation}
where $U_{ei}$ are electron-neutrino matrix elements of the Pontecovo-Maki-Nakagawa-Sakata (PMNS) mixing matrix, and $i$ sums over the three neutrino mass states, which have masses $m_{1}$, $m_{2}$, and $m_{3}$.  While the phenomenon of neutrino oscillations in vacuum indicates that all three mass states cannot be zero, it still allows for the possibility that  lightest mass state may be massless.

Less obvious is that there are also arguments for a lower, non-zero, bound on $m_{\nu_{e}}$ \cite{Fischbach AoP}, which follow from the analysis of many-body neutrino-exchange forces in compact objects such as neutron stars and white dwarfs.   It was shown in Ref.~\cite{Fischbach AoP} that if the lightest neutrino was massless, these many-body neutrino-exchange forces  within a neutron star would lead to an unphysically large energy density,  unless suppressed by some mechanism.  An example would be a  a non-zero neutrino mass which would limit the range of the forces.  Barring some other suppression mechanism, this argument implies that  the lightest neutrino mass  is   $\sim 0.4~{\rm eV}/c^{2}$.  Since this approximate limit is within reach of the anticipated target uncertainty of the KATRIN experiment, $\simeq 0.2~{\rm eV}/c^{2}$,  we will revisit here the lower-neutrino-mass-limit argument  in light of subsequent developments in neutrino physics since it was first proposed.  We will also quantify the remaining allowed parameter space for three SM neutrino mass states and examine the limits on sterile neutrino masses.

\begin{figure}[b]
\begin{center}
\includegraphics[height=1.5in]{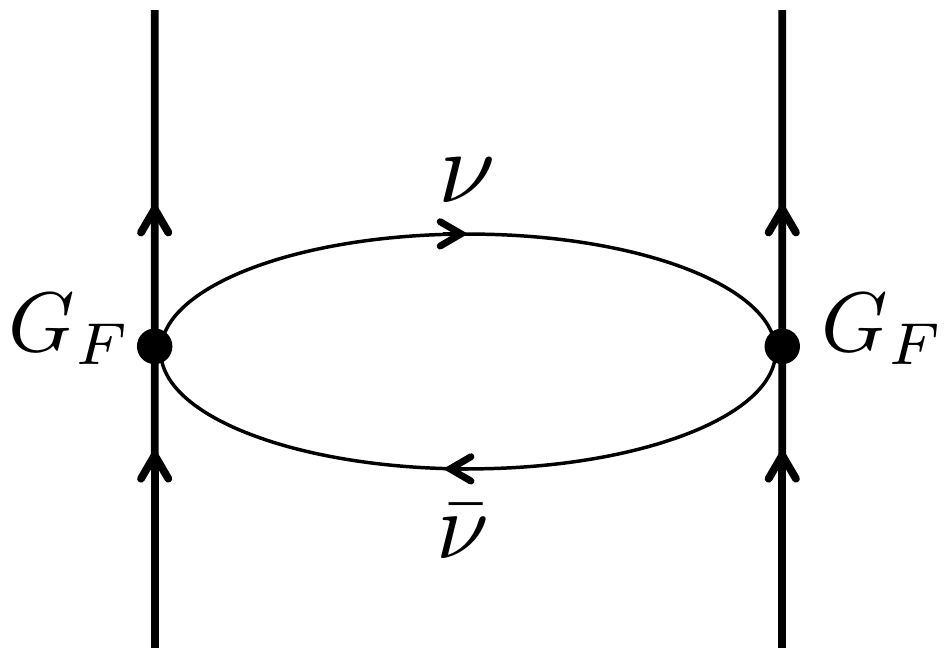}
\caption{Feynman diagram of neutrino-antineutrino exchange between fermions leading to Eq.~(\ref{FS potential}).  The thicker lines denote external  fermion propagators.}
\label{2NEP diagram}
\end{center}
\end{figure}

Within quantum field theory, macroscopically-ranged forces  naturally arise from the virtual exchange of light bosons, where the mass of the boson $m_{b}$ is related to the force range $\lambda$ by $\lambda = m_{b}^{-1}$.  (Throughout this paper,  we  use units such that $\hbar = c = 1$.)  Less well-known is  that long-range forces can also arise from the virtual exchange of light fermion-antifermion pairs.  Within the Standard Model, the only fermion candidates for a long-ranged fermion-antifermion interaction are neutrinos.  Feinberg and Sucher \cite{FS} were the first to apply modern methods to calculate the 2-neutrino exchange potential using the low-energy effective theory of the weak interaction.  For the diagram shown in Fig.~\ref{2NEP diagram}, they found for the exchange of massless neutrino-antineutrino pairs, that the interaction potential between two electrons separated by a distance $r$ is given by
\begin{equation}
V_{\nu,\bar{\nu}}(r) = \frac{G_{F}^{2}}{4\pi^{3}r^{5}},
\label{FS potential}
\end{equation}
where $G_{F}$ is the Fermi constant. The Feinberg-Sucher result was subsequently modified to include the neutral current interaction after it had been discovered \cite{FSA}.  Hsu and Sikivie \cite{Hsu} and  Segarra \cite{Segarra} also obtained these results using Feynman diagrammatic  and dispersion theory approaches, respectively.

All of preceding results for the 2-neutrino exchange potential (2NEP) assumed that the neutrinos were massless.  The next step, to generalize Eq.~(\ref{FS potential}) for the case of a neutrino with mass $m_{\nu}\neq 0$, was carried out by Fischbach~\cite{Fischbach AoP}, who found, after some simplification, 
\begin{equation}
V_{\nu,\bar{\nu}}^{D}(r) = \frac{G_{F}^{2}m_{\nu}^{3}}{4\pi^{3}r^{2}}K_{3}(2m_{\nu}r),
\label{massive D V}
\end{equation}
where $K_{n}(x)$ is the modified Bessel function.  This result, which applies to Dirac neutrinos, was also obtained shortly thereafter by Grifols {\em et al.} \cite{Grifols}, who also obtained an analogous result for for Majorana neutrinos:
\begin{equation}
V_{\nu,\bar{\nu}}^{M}(r) = \frac{G_{F}^{2}m_{\nu}^{3}}{8\pi^{3}r^{3}}K_{2}(2m_{\nu}r).
\label{massive M V}
\end{equation}
Both Eqs.~(\ref{massive D V}) and (\ref{massive M V}) reduce to Eq.~(\ref{FS potential}) in the massless limit; there is no distinction between Dirac and Majorana neutrinos in this case.   We also see from Eqs.~(\ref{massive D V}) and (\ref{massive M V}), and the large separation-dependence of the Bessel functions, that the effective range of the 2NEP is $\lambda \simeq (2m_{\nu})^{-1}$, which is what would be expected for the virtual exchange of two massive neutrinos.

So far, the 2NEPs given by Eqs.~(\ref{FS potential})--(\ref{massive M V}) assume only a single flavor of neutrino.  More recently, the 2NEP for Dirac neutrinos was extended to include the three SM flavors, mixing, and $CP$-violation \cite{LTK}, and evaluated for lepton-lepton, lepton-nucleon, and nucleon-nucleon interactions.  The simplest case, which is relevant here, is the 2NEP between two nucleons N$_{1}$ and N$_{2}$, which is given by
\begin{equation}
V_{\rm N_{1},N_{2}}(r) = \frac{G_{F}^{2}g_{V,1}^{{\rm N}_{1}}g_{V,2}^{{\rm N}_{2}}}{4\pi^{3}r^{2}}\sum_{i= 1}^{3}m_{i}^{3}K_{3}(2m_{i}r),
\label{N-N V}
\end{equation}
where the sum is over the neutrino mass states.   Here, the vector neutral current couplings to protons and neutrons are $g_{V}^{p} =1/2 - 2 \sin^{2}\theta_{W}$ and $g_{V}^{n} = -1/2$, respectively. $\theta_{W}$ is the Weinberg angle.
In the limit $r \ll m_{i}^{-1}$ for all $i$, this reduces to
\begin{equation}
 V_{\rm N_{1},N_{2}}(r) \simeq  \frac{3G_{F}^{2}g_{V,1}^{{\rm N}_{1}}g_{V,2}^{{\rm N}_{2}}}{4\pi^{3}r^{5}}.
\label{massless N-N V}
\end{equation}
This problem was also considered by Lusignoli and Petrarca \cite{LP}.    Recently, several authors have been exploring the short distance behavior of the 2NEP \cite{Xu}, and  the consequences of the 2NEP in atomic physics \cite{Stadnik,Asaka,Ghosh,Munro-Laylim}, the distinction between Dirac and Majorana neutrinos \cite{Segarra}, tests of the weak equivalence principle \cite{Fischbach WEP}, physics beyond the Standard Model \cite{Bolton}, and interactions with dark matter \cite{Orlofsky,Coy}.

\begin{figure}[b]
\begin{center}
\includegraphics[width=2in]{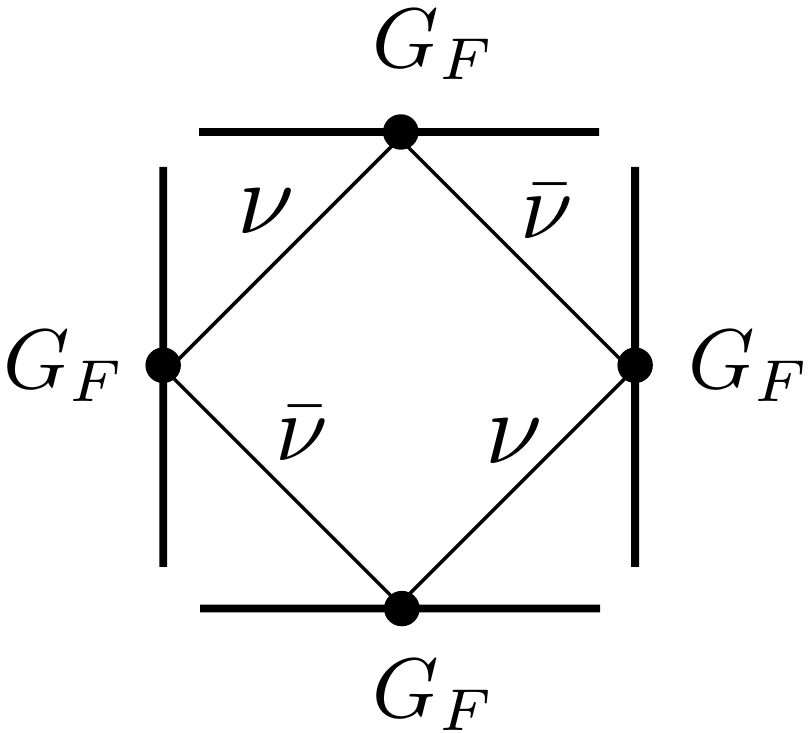}
\caption{A diagram illustrating the exchange of neutrino-antineutrino pairs that will give rise to 4-body potential that is not a product of 2-body potentials.  The thick lines represent external fermion propagators.  Calculating the total potential energy requires including all possible permutations of such diagrams.}
\label{4-body diagram}
\end{center}
\end{figure}

In the low-energy Fermi effective theory of the weak interaction with Dirac neutrinos, a fermion couples to neutrino-antineutrino pairs, as shown  in Fig.~\ref{2NEP diagram}.  This has the consequence that that multi-body interactions naturally arise when fermions do not exchange the same neutrino-antineutrino pairs \cite{Hartle,Fischbach AoP}.  As in the example shown in Fig.~\ref{4-body diagram}, an even number of fermions can exchange neutrino-antineutrino pairs that will give rise to a multi-body potential energy that is not simply a product of the 2-body 2NEP.

In Ref.~\cite{Fischbach AoP}, it was shown that the binding energy of a spherical body of radius $R$ consisting of $N$ neutrons due to the multi-body neutrino exchange potential is given by
\begin{equation}
W = W^{(2)} +  \sum^{N}_{k = 4,6,8, \ldots} W^{(k)},
\end{equation}
where $W^{(2)}$ is the contribution from the 2NEP, and $W^{(k)}$ is the $k$-body contribution.  For $k>2$ with a massless neutrino state,
\begin{equation}
W^{(k)} \sim \frac{1}{k!} \frac{1}{R}\left(\frac{G_{F}N}{R^{2}}\right)^{k}.
\label{Wk}
\end{equation}
For ordinary nuclei, the contribution due to the 2NEP, $W^{(2)}$, dominates and contributes a fraction $\sim 10^{-17}$ to the total rest energy of a nucleus \cite{Fischbach WEP}. However,  $W^{(k)}$ grows catastrophically for large, dense bodies, such as neutron stars, and quickly becomes much larger than the rest energy of the star itself.  While the other long-range interactions of the Standard Model (SM), gravity and electromagnetism, also lead to multi-body potentials, only a massless neutrino multi-body interaction has the essential characteristics which can lead to such dramatic effects in macroscopic bodies.  These include  long-range, and  large coupling strength, in bodies with a large bulk matter ``charge.'' 

The catastrophic effects arising from the massless multi-body neutrino interaction are most naturally avoided if the lightest neutrino mass state had a mass $m_{\nu,{\rm min}}$ sufficiently large such that a  neutrino in a neutron star only interacts with a sufficiently small subvolume of the star.  In Ref.~\cite{Fischbach AoP}, it was shown that this leads to a limit on the smallest neutrino mass state 
\begin{equation}
	\label{limit 1996}
m_{\nu,{\rm min}} \gtrsim \frac{\sqrt{2}G_{F}\rho}{6e^3} = \mbox{0.4 eV,}
\end{equation}
where $\rho$ is the number density of the neutron star.  The combination $G_{F}\rho$, which appears in the Mikheyev-Smirnov-Wolfenstein (MSW) mechanism \cite{Mikheyev,Wolfenstein} for neutrino oscillations in matter, is the only  quantity with dimensions of mass that can be formed by the dynamical variables of the system.

Alternatives to massive neutrinos for eliminating the neutrino binding energy catastrophe were considered in Ref.~\cite{Fischbach AoP}, including deviations from SM couplings of neutrinos to neutrons, breakdown of the perturbation theory expansion, and modifications of the interaction due to the neutron star medium.  However, all these were shown to be insufficient.  A number of authors have attempted to criticize a perturbation calculation of the neutron self-energy \cite{Abada 1996, Abada 1998,Kachelriess,Arafune}.  Other authors have argued that that the degenerate low-energy sea of neutrinos within a neutron would effectively screen the neutrino interactions \cite{Smirnov,Kiers}, but this was shown not be be the case \cite{Fischbach comment}.

\begin{figure}
\begin{center}
		\includegraphics[width=3.5in]{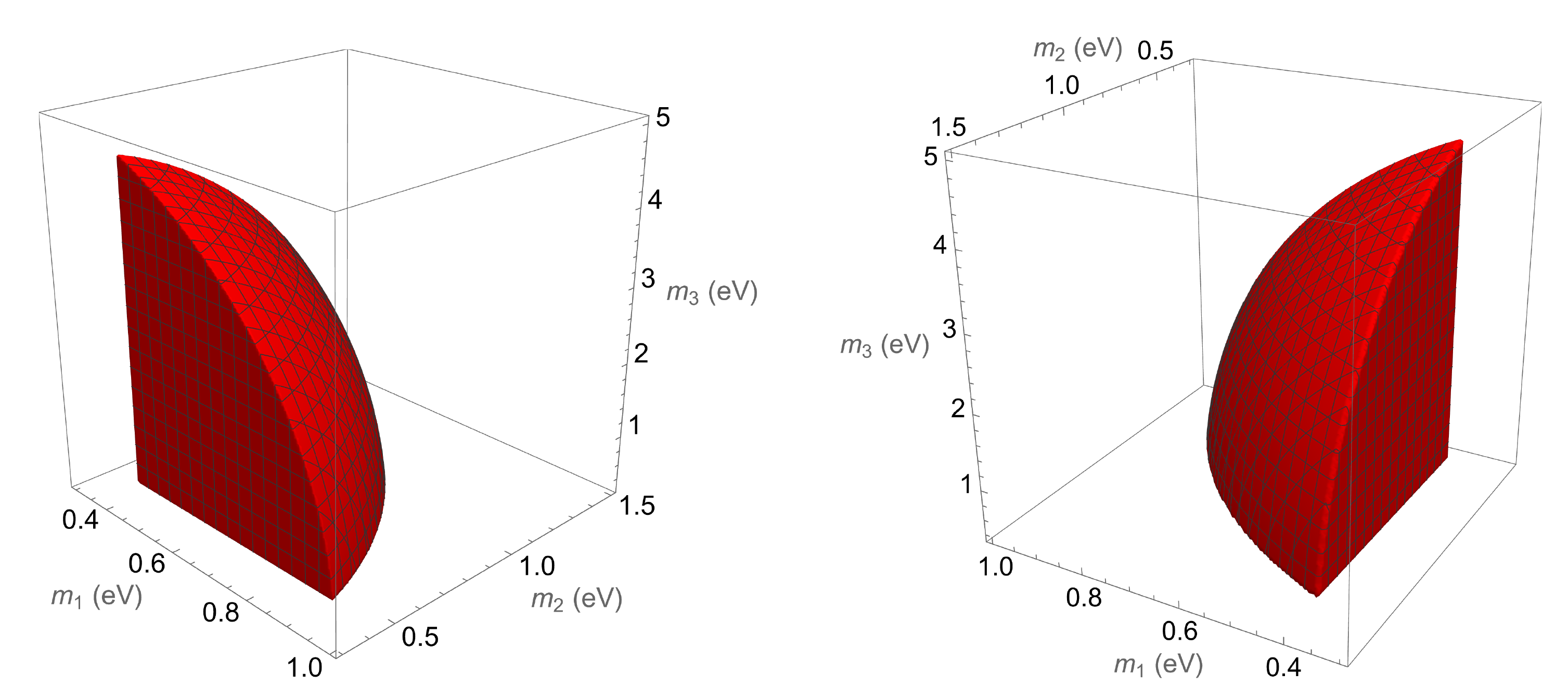}
		\end{center}
	\caption{Graphs showing the remaining allowed region in neutrino mass space when the KATRIN limit $m_{\nu_{e}}< 0.8~{\rm eV}$ is combined with the lower  neutrino mass eigenstate limit $m_i \gtrsim 0.4$ eV.}
	\label{neutrino corner}
\end{figure}

Combining the limit in Eq.~(\ref{limit 1996}), and latest limit for the effective electron anti-neutrino mass from KATRIN \cite{KATRIN} $m_{\nu_{e}} < 0.8 \ \mbox{eV}$,
we can visualize the remaining allowed region of three neutrino mass eigenstates in the mass space as the region bounded between a curve surface arising from the mathematical expression of $m_{\nu_{e}}$ and a corner stemming from our neutrino mass eigenstate limit (Fig.~\ref{neutrino corner}). 

\begin{figure}[b]
	\begin{center}
		\includegraphics[width=3.5in]{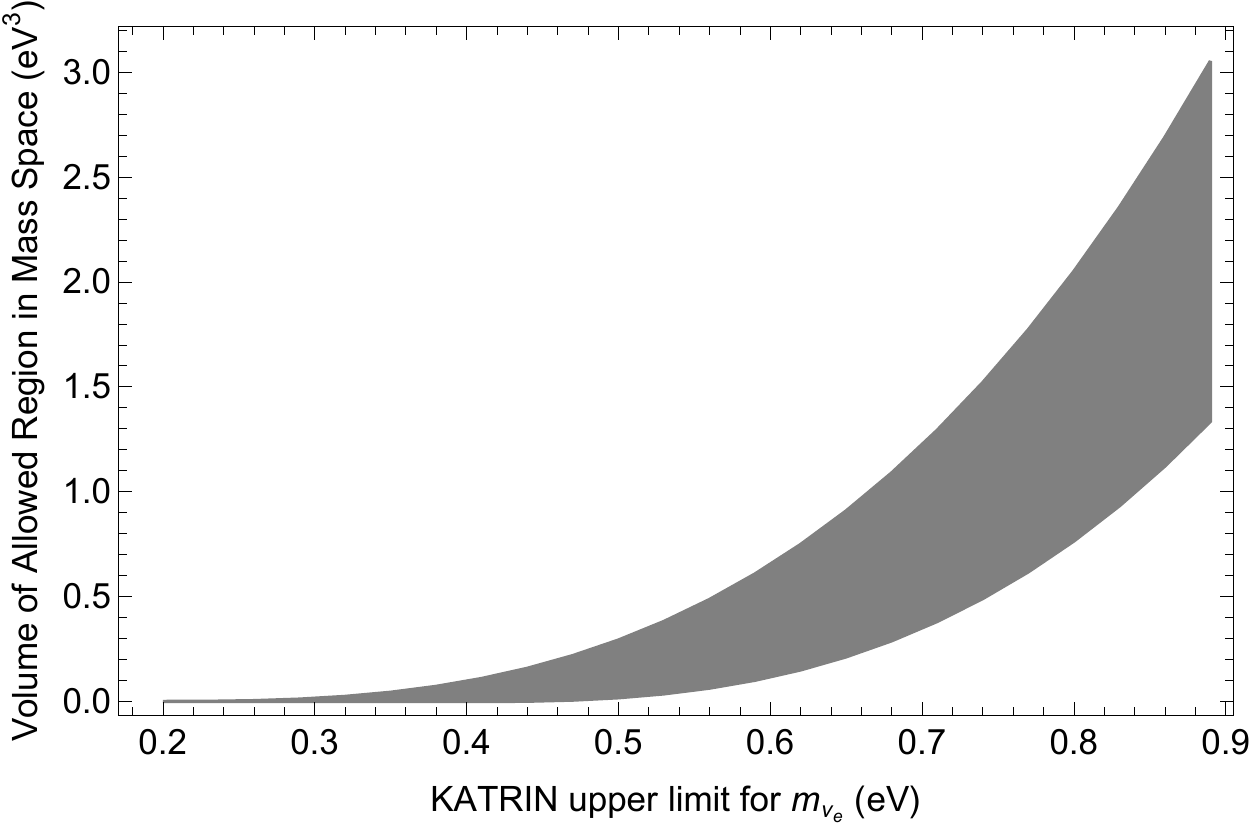}
		\end{center}
		\caption{Graph shows how the volume of allowed region in neutrino mass space varies with the effective electron anti-neutrino mass limit reached by KATRIN. The uncertainty band is due to the lower limit established by our calculation containing assumptions and variables in modeling physics of neutron stars. Here, this lower limit is assumed to vary from 0.2--0.4~eV.}
		\label{volume of allowed mass phase space}

\end{figure}

%{\color{red} 
The particular limit in Eq.~(\ref{limit 1996}) is derived assuming neutron star parameters from the Hulse-Taylor binary system PSR 1913+16 \cite{Hulse Taylor ,Taylor Weisberg,Will}. Since many variables and assumptions are involved in modeling the physics of neutron stars, we expect the lower bound of neutrino mass to vary. This is reflected in Fig.~\ref{volume of allowed mass phase space}, where there is an uncertainty band for the volume of allowed region in neutrino mass space. In particular, we assume that the lower limit of neutrino mass may vary between 0.2--0.4~eV. We see that this volume falls rapidly from the current values 0.7--2.0~eV$^3$ to less than 0.3~eV$^3$ once KATRIN reaches the very attainable limit of $m_{\nu_{e}} < 0.5$ eV.

The fundamental idea behind using virtual-exchange interaction energy is to explore all dynamical degrees of freedom participating in the interaction. For the neutrino sector, the mass eigenstates are the fundamental mediators contributing to our multi-body neutrino exchange potential. Currently, we expect neutrinos to have three mass eigenstates corresponding to the three known neutrino flavors from the weak interaction. There are indeed both theoretical and experimental motivations for the existence of sterile neutrino flavors, necessitating extra neutrino mass eigenstates \cite{Giunti2,Dasgupta}. While it is certainly true that the exact flavor mixing mechanism depends on specific Beyond Standard Model physics, within our framework we can parameterize neutrino mixing in a minimal $3+1$ model generically as
\begin{equation}
	\left(\begin{array}{c}
		\nu_{e} \\
		\nu_{\mu} \\
		\nu_{\tau} \\
		\nu_{s}
	\end{array}\right)=\left(\begin{array}{cccc}
		U_{e 1} & U_{e 2} & U_{e 3} & U_{e 4} \\
		U_{\mu 1} & U_{\mu 2} & U_{\mu 3} & U_{\mu 4} \\
		U_{\tau 1} & U_{\tau 2} & U_{\tau 3} & U_{\tau 4} \\
		U_{s 1} & U_{s 2} & U_{s 3} & U_{s 4}
	\end{array}\right)\left(\begin{array}{c}
		\nu_{1} \\
		\nu_{2} \\
		\nu_{3} \\
		\nu_{4}
	\end{array}\right),
\end{equation} 
where $e$, $\mu$ and $\tau$ are the SM flavors, $s$ is the supposed sterile flavor and $\nu_i$, with $i = 1, 2, 3$, and 4, are the mass-eigenstate fields. Assuming that the sterile flavor $s$ does not couple to neutrons via the weak interaction, the neutrino propagators of the active flavors inside loop integrals become
\begin{equation}
	S_\nu (E,\vec{r}_1, \vec{r}_2) = \sum_{f=e,\mu,\tau} \sum_{i=1}^{4} | U_{fi} |^2 S_i(E,\vec{r}_{1}, \vec{r}_{2}),
\end{equation} 
where $ S_i(E,\vec{r}_1, \vec{r}_2)$ is the propagator of the mass eigenstate $m_i$. This implies that the additional mass state $m_4$ also contributes to the multi-body neutrino exchange potential $W$. Furthermore, one important observation is that these additional contributions are always positive since $| U_{fi} |^2$ is non-negative. Therefore, in order to avoid the catastrophic effects from $W$, it is crucial for contributions from each mass eigenstate to be suppressed individually.

\begin{figure}[t]
	\begin{center}
		\includegraphics[width=3.5in]{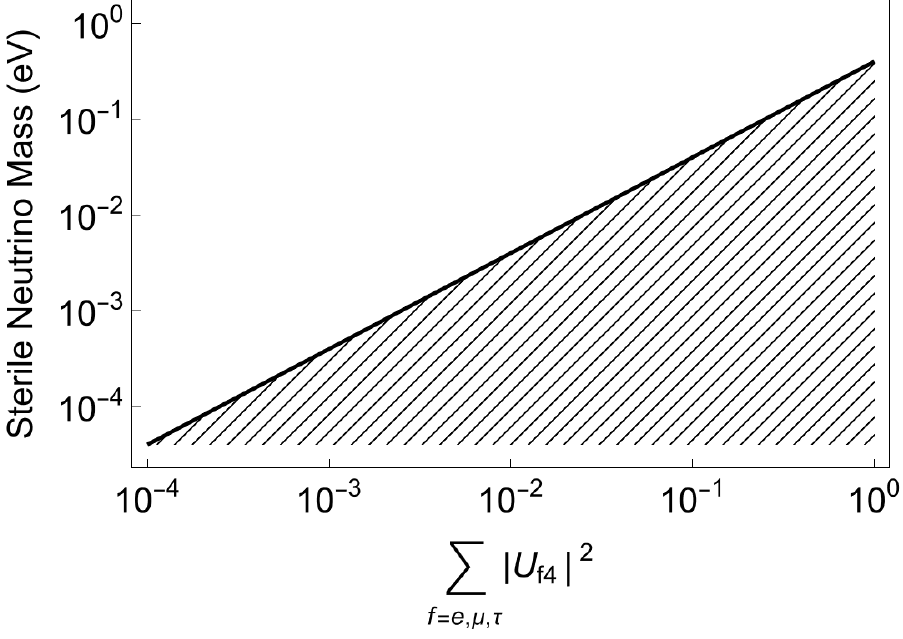}
			\end{center}
		\caption{Limit on the mass of the 4th-neutrino mass eigenstate $m_4$ in terms of the sum of mixing matrix elements between the 3 active neutrino active flavor and the 4th-neutrino mass eigenstate, assuming that the lightest active neutrino mass state is $m_{\nu,{\rm min}} =0.4$~eV. The hatched filling indicates the excluded region.  However, as described in the text, this constrain only applies to sterile neutrinos with a mass less than the lightest SM neutrino mass state.}
		\label{sterile neutrino plot}

\end{figure}

\begin{figure}[b]
	\begin{center}
		\includegraphics[width=3.5in]{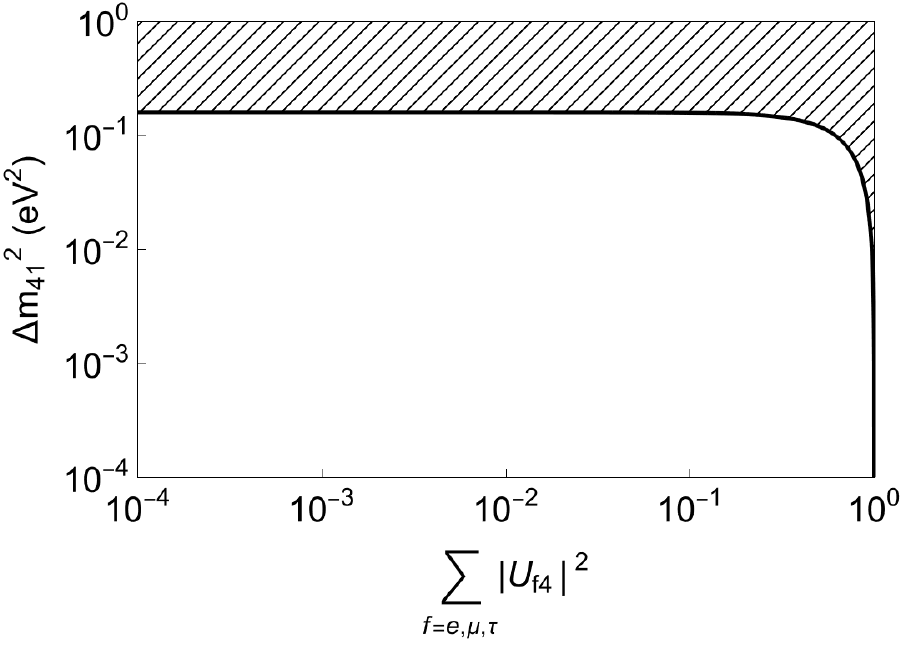}
			\end{center}
		\caption{Limit for mass-squared difference $\Delta m_{14}^2$ and the sum of mixing matrix elements between the 3 active neutrino active flavor and the 4th-neutrino mass eigenstate, assuming that the lightest active neutrino mass state is $m_1 = 0.4$~eV and $m_4 < m_1$. Hatch filling indicates excluded region.}
		\label{sterile neutrino mass difference plot}

\end{figure}

With these modifications, it is possible  to arrive at the lower limit for the mass of the 4th neutrino mass eigenstate given by
\begin{equation}
	m_4 \gtrsim \mbox{0.4 eV} \!\!\!\sum_{f=e,\mu,\tau} | U_{f4} |^2.
	\label{m4 limit}
\end{equation} 
Since the neutron couples to all three SM neutrino flavors, we obtain limits on $m_{4}$ in terms of the sum of all three mixing matrix elements squared between the SM neutrino flavors and the 4th-neutrino mass eigenstates, which is shown in Fig.~\ref{sterile neutrino plot}. This is a distinct feature of our approach compared to current oscillation experiments where the obtained limits involve only mixing matrix elements of flavors in production and detection of the active neutrino flavors \cite{Giunti2,Dasgupta,Diaz}. Using Eq.~(\ref{m4 limit}), we can obtain the limit for the typical mass-squared difference $\Delta m_{14}^2$ between the sterile and the lightest active mass state as shown in Fig.~\ref{sterile neutrino mass difference plot}. However, such limit on $\Delta m_{14}^2$ is only possible if the sterile state is lighter than the lightest active neutrino state $m_1$, because the sum of the mixing matrix element squared appeared in these limits is always less than unity due to a property of the unitary mixing matrix.

%An interesting observation is that since neutron couples equally all neutrino flavors, the Dirac CP in the 3-neutrino PMNS matrix is not observable. Once, we add the 4th-neutrino, 

In summary, we have re-examined and updated the arguments supporting a lower-bound on neutrino masses based on the application of the multi-body neutrino interactions and their consequences on astrophysical bodies.  We then combined the lower-bound with the newly published upper-bound on $m_{\nu_{e}}$ from the KATRIN experiment to determine the remaining available parameter space for the neutrino masses.  If we use reasonable estimates for the properties of neutron stars and the expected final sensitivity on $m_{\nu_{e}}$ that the KATRIN collaboration hopes to achieve, the available parameter space could close completely within the next few years.  Finally, we extended the lower-bound argument when a sterile neutrino flavor is present.  We find that the neutrino mixing allows one to also obtain a lower bound on a 4th neutrino mass eigenstate.

%\acknowledgments

\vspace{12pt}

The authors wish to thank Magnus Schl\"osser for a very helpful communication with respect to future KATRIN sensitivity to $m_{\nu_{e}}$. QLT thanks Radovan Dermisek for useful discussions.

\end{document}